\documentclass[12pt, centerh1]{article}

\usepackage{graphicx,colonequals,bm}
\usepackage{natbib}
\usepackage{url}
\usepackage{amsmath}
\usepackage{amsfonts}
\usepackage{amssymb}
\usepackage{url}
\usepackage{color}
\usepackage{subfigure}
\newcommand{\bb}{\mathbf{b}}
\newcommand{\bx}{\mathbf{x}}
\newcommand{\by}{\mathbf{y}}

\newcommand{\bX}{\mathbf{X}}

\newcommand{\cL}{\mathcal{L}}

\newcommand{\cX}{\mathcal{X}}

\newcommand{\EE}{\mathbb{E}}

\newcommand{\bmu}{\mbox{\boldmath $\mu$}}

\newcommand{\bpsi}{\mbox{\boldmath $\psi$}}

\newcommand{\bbeta}{\mbox{\boldmath $\beta$}}
\newcommand{\bchi}{\mbox{\boldmath $\chi$}}

\newcommand{\btheta}{\mbox{\boldmath $\theta$}}

\newcommand{\bpi}{\mbox{\boldmath $\pi$}}

\newcommand{\bSigma}{\mbox{\boldmath $\Sigma$}}

\begin{document}

\title{Accelerated Failure Time Models for Competing Risks in a Cluster Weighted Modelling Framework}
\author{Utkarsh J.\ Dang and Paul D.\ McNicholas\thanks{Department of Mathematics \& Statistics, University of Guelph, Guelph, Ontario, N1G 2W1, Canada. E-mail: udang@uoguelph.ca.}}
\date{Department of Mathematics \& Statistics, University of Guelph}
\maketitle

\begin{abstract}
A novel approach for dealing with censored competing risks regression data is proposed. This is implemented by a mixture of accelerated failure time (AFT) models for a competing risks scenario within a cluster-weighted modelling (CWM) framework. Specifically, we make use of the log-normal AFT model here but any commonly used AFT model can be utilized. The alternating expectation conditional maximization algorithm (AECM) is used for parameter estimation and bootstrapping for standard error estimation. Finally, we present our results on some simulated and real competing risks data.
\end{abstract}

\section{Introduction}\label{sec:introduction}

\section{Introduction}
\label{intro}

Survival techniques are useful in situations where regular regression procedures are inadequate. For instance, where the probability of survival past a certain time is of more interest than the expected time of survival, survival techniques are more appropriate. The same holds true when there are censored data. Censored data may arise due to partial follow-up (until a specific time), loss to follow-up, or a subject dropping out of a study. Attrition is quite common in medical studies where a subject is followed to a particular event of interest. In such cases, the status of these subjects with respect to the event of interest is only known up to a particular time point.

Competing risks data often arise in survival and reliability analyses. These multiple cause data represent time to failure (occurrence of event of interest) and can be due to a host of causes. In a cancer study, for example, a significant number of deaths due to causes other than cancer may be expected post-treatment because the average age of cancer patients is high \citep{boag1949}.

Some existing methods for dealing with failure time data include parametric mixture models based on failure time densities \citep{farewell1986}, proportional hazards models \citep{cox1972}, cause-specific hazard functions \citep{prentice1978}, accelerated failure time models \citep{yamaguchi1992}, first hitting time models \citep{balka2009}, frailty models \citep{price2001}, and so on. A classical mixture model for competing risks adopted a proportional hazards model with piecewise constant baseline hazards \citep{larson1985}. A traditional approach to competing risks is the latent failure time approach \citep{david1978} which makes untestable assumptions about the independence of the competing risks; this is unlike the mixture approach \citep{mclachlan2000}.

We present a novel method to account for censored competing risks data using a mixture of accelerated failure time models in a cluster-weighted modelling (CWM) framework. This approach can be called the cluster-weighted accelerated failure time (CWAFT) model. Section \ref{cwm} gives an introduction to cluster-weighted modelling. Section \ref{methods} briefly goes over accelerated failure time models and the mathematical details involved for CWAFT models. In Section \ref{results}, we present our results on some simulated and real censored data, where the time to failure and cause of failure were only recorded for some of the observations. Finally, we conclude with a discussion and some ideas for further research.

\section{Cluster-weighted modelling}
\label{cwm}
\cite{mclachlan2000} give a good introduction to mixture models for the analysis of failure time data. In survival analysis, mixture models can be defined in terms of the density function $f(t)$ or in terms of the survival function $S(t)$ of $T$ where $T$, is the (non-negative, continuous) random variable representing the time to the event of interest. Hence, $f(t)=\sum^G_{g=1}\pi_{g}f_{g}(t)$ or $S(t)=\sum^G_{g=1}\pi_{g}S_{g}(t)$, where $f_{g}(t)$ and $S_{g}(t)$ are the $g$th component density and the $g$th component survival function, respectively.

\cite{ingrassia2012} introduced CWM in a more general statistical context. This methodology allowed for the decomposition and modelling of the joint probability of a response variable and a set of explanatory variables. Let $(\bX,Y)$ be the pair of random vector $\bX$ and random variable $Y$ defined on $\Omega$ with joint probability distribution $p(\bx,y)$. Here, the response variable $Y$ has values in $\mathcal{Y} \subseteq \mathbb{R}$ and the explanatory variable $\bX$ is a $d$-dimensional vector with values in $\cX \subseteq \mathbb{R}^d$. If $\Omega$ can be partitioned into $G$ disjoint groups, such that $\Omega= \Omega_1 \cup \cdots \cup \Omega_G$, the joint probability $p(\bx,y)$ can be decomposed as
\begin{equation*}
p(\bx,y| \btheta)=\sum^{G}_{g=1} p(y|\bx,\Omega_g)\, p(\bx|\Omega_g)\, \pi_g,
\end{equation*}
where $p(y|\bx,\Omega_g)$ is the conditional density of $Y$ given $\bx$ and  $\Omega_g$, $p(\bx|\Omega_g)$ is the probability density of $\bX$ given $\Omega_g$, , and $\pi_g=p(\Omega_g)$ is the mixing weight, where $\pi_{g}>0$ ($g = 1,\ldots,G$) and $\sum^G_{g=1}\pi_{g}=1$. $\btheta$ denotes the set of all parameters. Here, $\pi_g=p(\Omega_g)$ is the mixing weight of $\Omega_g$, where $\pi_{g}>0$ ($g = 1,\ldots,G$) and $\sum^G_{g=1}\pi_{g}=1$. The mixing proportions $\pi_g$ in survival mixture models are often assumed to follow a logistic model of the covariates \citep{farewell1986, ng2003}. Using a CWM framework, the model presented in this paper is not necessarily bound by this restriction. Here, $p(y|\bx,\Omega_g)$ can be viewed as weighted by both the marginal $p(\bx|\Omega_g)$ and unrestricted mixing weights $\pi_g$.

\section{Methodology}
\label{methods}
\subsection{Log-normal accelerated failure time model}
\label{ss:aft}
Accelerated failure time (AFT) models are a popular parametric regression alternative to proportional hazards regression. Some common AFT models employ the log-normal, log-logistic, and Weibull distributions \citep{lawless2003}. $Y=\log T$ is modeled analogous to classical regression. For a log-normal distribution,
\begin{equation*}
S_{T}(t)=1-\Phi\left(\frac{\log{t}-\mu}{\sigma}\right),
\end{equation*}
where $\mu$ is the mean and $\sigma$ is the standard deviation of $\log{T}$. This is $S_{W}(\frac{y-\mu}{\sigma})$ where $S_{W}(w)$ is the survival function of the standard Gaussian distribution. A regression model can be constructed for $Y$ given $\bx$ with
\begin{equation*}
S(y|\mu,\sigma)=S_{0}\left(\frac{y-\mu(\bx)}{\sigma}\right).
\end{equation*}

With covariates, the linear model representation is $y=\log{t}=\mu(\bx)+\sigma w=b_{0}+\bb'\bx+\sigma w$, where $W$ is the error distribution (log-scale form) and $\bx$ provides the covariates.  If a standard Gaussian distribution is used for the error distribution, this results in a log-normal AFT model. 

The underlying density and survival function for $Y$ are
\begin{equation*}
S_{Y}(y)=1-\Phi\left(\frac{y-\mu}{\sigma}\right)
\end{equation*}
and
\begin{equation*}
f_{Y}(y)=\frac{1}{\sigma}\phi\left(\frac{y-\mu}{\sigma}\right)=\frac{1}{\sigma}f_{W}\left(\frac{y-\mu}{\sigma}\right).
\end{equation*}

Here, $\bX \sim N_{d}(\bmu,\bSigma)$ and because $Y= b_{0}+\bb'\bx_i +\sigma w$, where $W \sim N(0,1)$, $Y \sim N(b_{0}+\bb'\bx_i,\sigma^2)$.

\subsection{Inference}
\label{ss:inference}
We propose that $p(y|\mathbf{x},\Omega_{g})$ in the CWM joint likelihood decomposition can be modeled as a Gaussian distribution where the random variable $Y$ is $\log{T}$, where $T$, the time to event, follows a log-normal distribution. Recall that CWM decomposes the joint probability $p(\bx,y)$ as follows:
\begin{equation*}
p(\bx,y| \btheta)=\sum^{G}_{g=1} p(y|\bx,\Omega_g)\, p(\bx|\Omega_g)\, \pi_g. 
\end{equation*}
The observed data likelihood is then
\begin{align*}
L_o(\btheta| \bX, \by) &=  \prod_{i=C+1}^{N} \prod_{g=1}^G [f_{Y}(y_{ig}|\bx_i, \bchi_g) \phi_d(\bx_i| \bpsi_g) \pi_g]^{l_{ig}} \times   \\
& \qquad \prod_{i=1}^{C} \sum_{g=1}^G S_{Y}(y_{ig}|\bx_i, \bchi_g)\phi_d(\bx_i| \bpsi_g) \pi_g,
\end{align*}
where $l_{ig}$ is 1 if observation $i$ is known to have failed from cause $g$ and 0 otherwise, $\bchi_g=(\bbeta_g,\sigma^2_g)$, and $\bpsi_g=(\bmu_g, \bSigma_g)$. $C$ and $N$ are the number of censored and total observations, respectively. 

The expectation-maximization framework is a popular technique for parameter estimation \citep{dempster1977}. This involves maximizing the expected complete-data log-likelihood. We use the alternating expectation conditional maximization (AECM) algorithm \citep{meng1997}, which is a variation of the EM algorithm that allows for different complete data at different stages. The EM algorithm was coded in R \citep{R2012}. Let ${(\bx_1,y_1), \ldots, (\bx_N,y_N)}$ be a sample of $N$ independent observations. Then, the corresponding complete-data likelihood function can be written in the form
\begin{equation*}
L_c(\btheta| \bX, \by) = \prod_{i=1}^N \prod_{g=1}^G [f_{Y}(y_{ig}|\bx_i, \bchi_g) \phi_d(\bx_i| \bpsi_g) \pi_g]^{z_{ig}}.
\end{equation*}

Here, two latent variables are introduced. $z_{ig}=1$ if $(\bx_i,y_i)$ comes from the $g$th population and $z_{ig}=0$ otherwise. This corresponds to the conditional probability that individual $i$ will eventually fail from risk $g$, conditional on no failure having occured by time $t$. Also, $y_{ig}=\log{t_{ig}}$ represents the uncensored failure time conditional on the $n$th individual failing from the $g$th cause ($g$th component). For uncensored individuals, $t_{ig}$ is observed; that is, it is merely $t_{i}$. As an illustration, consider $g=2$ and assume that $g=1$ corresponds to the competing risk and $g=2$ corresponds to the cause of interest. Then, $t_{i1}$ represents the failure time associated with the censored time $t_{i}$ conditional on the $i$th individual failing from the competing cause.
Using the accelerated failure time framework, the complete-data log-likelihood can be decomposed as
\begin{align*}
\cL_c (\btheta| \bX, \by) & = \log L_c(\btheta| \bX, \by) \\
& =\sum_{i=1}^N \sum_{g=1}^G [ z_{ig} \log\left(\frac{1}{\sigma_{g}} f_{W}(y_{ig}|\bx_i, \bchi_g)\right) + z_{ig} \log  \phi_d(\bx_i| \bpsi_g) + z_{ig} \log \pi_g ] \\
& = \cL_{1c} (\bchi) + \cL_{2c} (\bpsi) + \cL_{3c} (\bpi), 
\end{align*}
where 
\begin{align*}
\cL_{1c} (\bchi) & =\frac{1}{2} \sum_{i=1}^N \sum_{g=1}^G  z_{ig} \left[- \log 2\pi - \log \sigma_{g}^{2} - \frac{[y_{ig} - (\bb'_g\bx_i + b_{0g})]^2}{\sigma_{g}^{2}}\right] \\ 
\cL_{2c} (\bpsi) & =\frac{1}{2} \sum_{i=1}^N \sum_{g=1}^G  z_{ig} \left[- p \log 2\pi - \log |\bSigma_g| - (\bx_i-\bmu_g)' \bSigma_g^{-1} (\bx_i-\bmu_g)\right] \label{L2c} \\ 
\cL_{3c} (\bpi) & = \sum_{i=1}^N \sum_{g=1}^G  z_{ig} [\log \pi_g] .
\end{align*}

Given the parameters $\pi_{g}^{(k)}$, $\bmu_{g}^{(k)}$, $\bSigma_{g}^{(k)}$, $\bbeta_{g}^{(k)}$, $\sigma_{g}^{2(k)}$ at the $k$th iteration, the expected complete-data log-likelihood is
\begin{align*}
Q(\btheta,\btheta^{(k)}) & = \EE_{\theta^{(k)}} \{ \cL_c(\btheta|\bX, \by)\}  \notag \\
& = \sum_{i=1}^N \sum_{g=1}^G  \EE_{\theta^{(k)}} \{z_{ig}|\bx_i, y_i\} [ Q_{1}(\bchi_g|\btheta^{(k)}) + Q_{2}(\bpsi_g|\btheta^{(k)}) + \log \pi_g^{(k)}]  \notag \\  
& =\sum_{i=1}^N \sum_{g=1}^G  \tau_{ig}^{(k)} [ Q_{1}(\bchi_g|\btheta^{(k)}) + Q_{2}(\bpsi_g|\btheta^{(k)}) + \log \pi_g^{(k)}],  
\end{align*}
where
\begin{equation*} 
\tau_{ig}^{(k)} =  \EE_{\theta^{(k)}} \{z_{ig}|\bx_i, y_i\} = \frac{\pi_g^{(k)} S(y_i^{*}|\bx_i, \bbeta_g^{(k)},\sigma_g^{2(k)}) \phi_d(\bx_i| \bmu_g^{(k)}, \bSigma_g^{(k)})} {\sum_{j=1}^G \pi_j^{(k)} S(y_i^{*}|\bx_i, \bbeta_j^{(k)},\sigma_j^{2(k)}) \phi_d(\bx_i| \bmu_j^{(k)}, \bSigma_j^{(k)})} 
\end{equation*}
provides the current value on the $k$-iteration for the censored observations and
\begin{align*} 
Q_{1}(\bchi_g|\btheta^{(k)}) &= \frac{1}{2} \left[ -\log 2\pi - \log \sigma_{g}^{2} - \EE^{(k)}\left(\frac{[y_{ig} - (\bb'_g\bx_i + b_{0g})]^2}{\sigma_{g}^{2}}\right) \right], \\
Q_{2}(\bpsi_g|\btheta^{(k)}) & =  \frac{1}{2} \left[- p \log 2\pi - \log |\bSigma_g| - (\bx_i-\bmu_g)' \bSigma_g^{-1} (\bx_i-\bmu_g)\right].
\end{align*}

Here, for an observed failure time, $\EE^{(k)}\left(y_{ig}\right)=y_{i}$. For right censored times $t_i^*$ (associated response variables are $y_i^*$), following \cite{schmee1979} we observe that the conditional probability density function of an observation censored to the right is that of the random variable $t$ truncated to the left at $t^*$. Hence,
\begin{equation*}
f(y_{ig}^*)=\frac{f(y_{i})}{1-F(y_{i}^*)},  
\end{equation*}
where $y_{ig}^*=\log{t_{ig}^*}$ in which $t_{ig}^*$ is the lifetime that would have been observed if $t_{i}^*$ has not been censored; that is, if it had been observed. Then, the conditional expected value $\EE^{(k)}(y_{ig})$ for a right censored time is
\begin{equation*}
\EE^{(k)}(y_{ig})=\EE^{(k)}(y_{ig}|\mu^{(k)}_{g(y)},\sigma^{(k)}_g, y_{i}^*)=\mu^{(k)}_{g(y)}+\sigma^{(k)}_g\frac{\phi\left(\frac{y_{i}^*-\mu^{(k)}_{g(y)}}{\sigma^{(k)}_g}\right)}{1-\Phi\left(\frac{y_{i}^*-\mu^{(k)}_{g(y)}}{\sigma^{(k)}_g}\right)},  
\end{equation*}
where $\mu_{g(y)}=\bb'_g\bx_i + b_{0g}$. Note that this is also the mean of the truncated normal distribution (see Appendix \ref{AppEyderive}). Similarly, $\EE^{(k)}(y_{ig}^2)$ is (see Appendix \ref{AppEy2derive})
\begin{align*}
\EE^{(k)}(y_{ig}^2) &= \sigma_g^{2(k)}\left[\frac{\left(\frac{y^*-\mu^{(k)}_{g(y)}}{\sigma^{(k)}_g}\right)\phi{\left(\frac{y^*-\mu^{(k)}_{g(y)}}{\sigma^{(k)}_g}\right)} + 1 - \Phi\left(\frac{y^*-\mu^{(k)}_{g(y)}}{\sigma^{(k)}_g}\right)}{1-\Phi\left(\frac{y^*-\mu^{(k)}_{g(y)}}{\sigma^{(k)}_g}\right)}\right] +\\
& \qquad  2 \mu^{(k)}_{g(y)} \EE(y_{ig}) - \mu^{2(k)}_{g(y)}.   
\end{align*}

The M-step requires the maximization of the conditional expectation of the complete-data log-likelihood with respect to $\btheta$. The updates for the parameters $\pi_{g}^{(k+1)}$, $\bmu_g^{(k+1)}$, and $\bSigma_g^{(k+1)}$ can be found by taking the derivative with respect to the appropriate parameter
\begin{align*}
\hat{\pi}_{g}^{(k+1)} &= \frac{1}{N} \sum_{i=1}^N \tau_{ig}^{(k)}, \label{pi_k} \\
\hat{\bmu}_g^{(k+1)} &= \frac{\sum_{i=1}^N \tau_{ig}^{(k)}  \bx_i}{\sum_{i=1}^N \tau_{ig}^{(k)} }, \notag \\
\hat{\bSigma}_g^{(k+1)} &= \frac{\sum_{i=1}^N \tau_{ig}^{(k)}  (\bx_i - \bmu_g^{(k+1)})(\bx_i - \bmu_g^{(k+1)})'}{\sum_{i=1}^N \tau_{ig}^{(k)} }. \notag
\end{align*}
These closed form updates can be found in \cite{mclachlan2000}. 
The updates $\bb_{g}^{(k+1)},b_{0g}^{(k+1)}$, and $\sigma_{g}^{2(k+1)}$ can similarly be derived (Appendix \ref{mstepapp}):

\begin{align*}
\hat{b}_{0g}^{(k+1)} &= \frac{\sum_{i=1}^N \tau_{ig}^{(k)}\EE(y_{ig})}{\sum_{i=1}^N \tau_{ig}^{(k)}}-\bb_{g}^{'(k+1)}\frac{\sum_{i=1}^N \tau_{ig}^{(k)} \bx_i}{\sum_{i=1}^N \tau_{ig}^{(k)}},\\
\hat{\bb}_{g}^{'(k+1)} &= \left(\frac{\sum_{i=1}^N \tau_{ig}^{(k)} \EE(y_{ig}) \bx'_n}{\sum_{i=1}^N \tau_{ig}^{(k)}}-\frac{\sum_{i=1}^N \tau_{ig}^{(k)} \EE(y_{ig})}{\sum_{i=1}^N \tau_{ig}^{(k)}}\frac{\sum_{i=1}^N \tau_{ig}^{(k)} \bx'_n}{\sum_{i=1}^N \tau_{ig}^{(k)}}\right) \times  \\
& \qquad \left(\frac{\sum_{i=1}^N \tau_{ig}^{(k)} \bx_i\bx'_n}{\sum_{i=1}^N \tau_{ig}^{(k)}}-\frac{\sum_{i=1}^N \tau_{ig}^{(k)} \bx'_n}{\sum_{i=1}^N \tau_{ig}^{(k)}} \left(\frac{\sum_{i=1}^N \tau_{ig}^{(k)} \bx'_n}{\sum_{i=1}^N \tau_{ig}^{(k)}}\right)^{'}  \right)^{-1}, \\
\hat{\sigma}_{g}^{2(k+1)} &= \frac{\sum_{i=1}^N \tau_{ig}^{(k)} \left[\EE(y_{ig}^{2})-2\left(\bb_{g}'^{(k+1)}\bx_i+b_{0g}^{(k+1)}\right)\EE(y_{ig})+\left(\bb_{g}'^{(k+1)}\bx_i+b_{0g}^{(k+1)}\right)^2\right]} {\sum_{i=1}^N \tau_{ig}^{(k)}} .
\end{align*}

Aitken's stopping criterion was used to determine convergence of the algorithm. Aitken's acceleration at iteration $k$ can be calculated as 
\begin{equation*}
a^{(k)}=\frac{l^{(k+1)}-l^{(k)}}{l^{(k)}-l^{(k-1)}},
\end{equation*}
where $l^{(k-1)}$, $l^{(k)}$, and $l^{(k+1)}$ are the log-likelihood values from iterations $k-1$, $k$, and $k+1$, respectively. This can be used to compute the asymptotic estimate of the log-likelihood at iteration $k+1$ \citep{bohning1994}:
\begin{equation*}
l_{A}^{(k+1)}=l^{(k)}+\frac{1}{1-a^{(k)}}(l^{(k+1)}-l^{(k)}).
\end{equation*}
The EM algorithm is stopped when $l_A^{(k+1)}-l^{k+1}<\epsilon$ \citep{lindsay1995}. 

\section{Results}
\label{results}
We use overall survival functions and cumulative incidence function for judging model fits. The overall survival function is estimated in the fashion of \cite{larson1985} by averaging individual estimates of subjects, and compared to a nonparametric estimate by way of a Kaplan Meier curve \citep{kaplan1958}. This statistic does not utilize information on cause of death or covariates. The overall survival function is $\hat{G}(t)=\sum_{i=1}^N \frac{1}{N} \sum_{g=1}^G\hat{\pi}_g \hat{S}_g(t|\bx_i)$.

Since Kaplan-Meier (KM) curves are not appropriate for the event of interest or the competing event directly, cumulative incidence curves are used. The cumulative incidence function can be defined as the cause-specific failure probability up to a certain time point $t$. The cumulative incidence curves are also presented with fits to the Nelson-Aalen cumulative incidence estimator \citep{cmprsk2011}. Following \cite{ng2003}, the cumulative incidence function for the $g$th type of failure is $\pi_g(1-\hat{S}_g(t))$. This was estimated by $\pi_g\left(1-\sum_{i=1}^N \frac{1}{N} \hat{S}_g(t| \bx_i)\right)$.

The standard errors were calculated using a non-parametric bootstrap \citep{efron1979} that was adjusted to account for the competing risks structure as in \cite{ng1999}. As before, let $C$ be the number of censored observations. Then, let $N_g$ $(g=1, \ldots, G)$ be the number of observed failures due to the $g$th cause. Each bootstrap sample was obtained by sampling with replacement separately from each of the $G+1$ sets with the size of each bootstrap subsample equal to $N_g$ and $C$, respectively. The data $(\bx_i,y_i)$ were re-sampled 100 times independently. The standard deviation of the resulting bootstrap maximum likelihood estimates yield an approximation of the standard error of the estimates.

\subsection{Simulation}
\label{simul}
Note that we present results for a simulation with two covariates here for simplicity; however, the model is not bound by this restriction and seems to perform quite well in simulations with more than two covariates.

\subsubsection{Simulated Data}
\label{simulation}
The specifications for the Gaussian covariate vectors for group 1 were mean$=(0.5, 2.3)'$ and covariance $\bigl(\begin{smallmatrix}
0.05&0\\ 
0&0.15
\end{smallmatrix} \bigr)$. The specifications for the Gaussian covariate vectors for group 2 were mean$=(0.7, 1.8)'$ and covariance $\bigl(\begin{smallmatrix}
0.20&0\\ 
0&0.20
\end{smallmatrix} \bigr)$. The regression intercepts used for the groups were $b_{01}=2$ and $b_{02}=1.4$, respectively. The regression coefficients were $b_{11}=(1.3, 0.8)$ and $b_{12}=(1.4, 1.3)$, respectively, and the error $\epsilon \sim N(0,1)$.  Fifty observations were censored (type II right, noninformative) in total by subtracting normally distributed values from randomly selected observations from both groups.\\

\begin{table}
\caption{Parameter estimates and standard errors (rounded off to 2 decimals) for simulated data.} 
\label{parestsim2}
\begin{tabular}{p{0.8in}p{1.2in}p{1.2in}p{1.2in}p{1.2in}}
\hline
Parameter & Estimates (g=1) & Standard error (g=1) & Estimates (g=2) & Standard error (g=2)\\
  \hline
$\pi_g$ &  0.48 & 0.01 & 0.52  & 0.01\\ 
$\bmu_g$ & $\left(0.47, 2.30 \right)'$ & $\left(0.02, 0.04\right)'$ & $\left(0.68, 1.84 \right)'$ & $\left(0.06, 0.05\right)'$\\
$\bSigma_g$  & $\begin{pmatrix}
  0.05 & -0.01 \\
  -0.01 & 0.15 
 \end{pmatrix}$ & $\begin{pmatrix}
  0.01 & 0.01 \\
  0.01 & 0.03 
 \end{pmatrix}$ & $\begin{pmatrix}
  0.27 & -0.02 \\
  -0.02 & 0.22 
 \end{pmatrix}$ & $\begin{pmatrix}
  0.03 & 0.03 \\
  0.03 & 0.03 
 \end{pmatrix}$ \\
$b_{0g}$ & 2.05 & 0.64 & 2.64 & 0.50\\
$\bb_{g}$ &  $\left(0.99, 0.93\right)'$ & $\left(0.48, 0.25\right)'$ & $\left(1.35, 0.84\right)'$  &  $\left(0.24, 0.25 \right)'$\\
$\sigma^2_{g}$ & 0.90 & 0.04 & 1.13 & 0.11\\
   \hline
\end{tabular}
\end{table}

\begin{figure}
\includegraphics{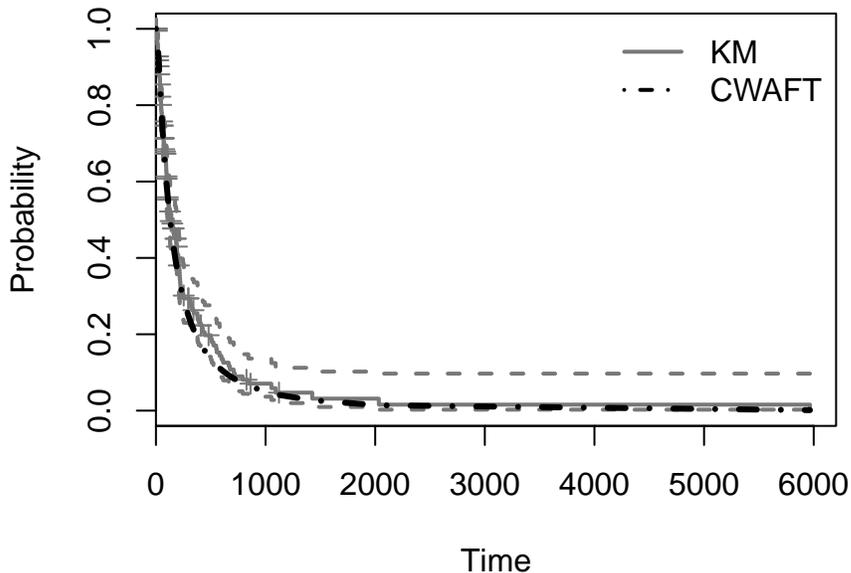}
\caption{Estimated overall survival function overlaid on the Kaplan-Meier estimator and its associated confidence intervals for simulation 2.}
\label{sim2osf}
\end{figure}

\begin{figure}
\includegraphics{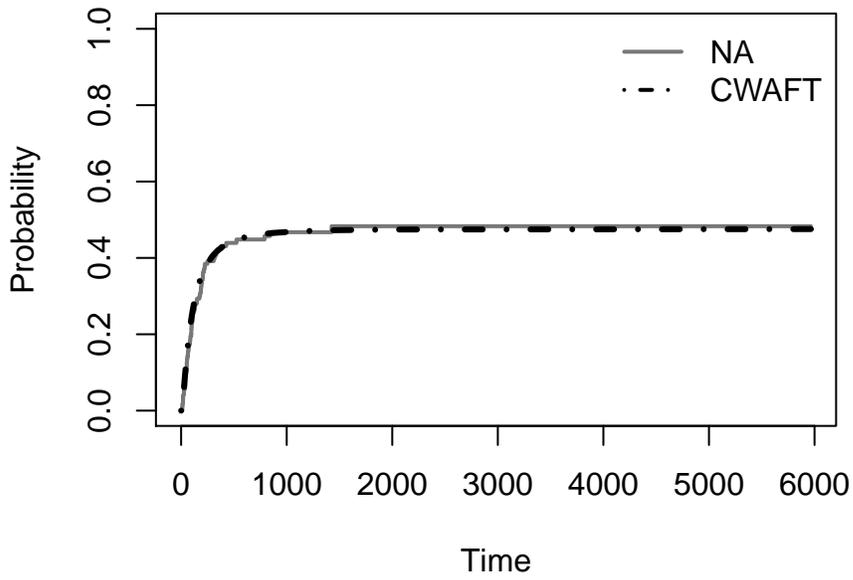}\\
\includegraphics{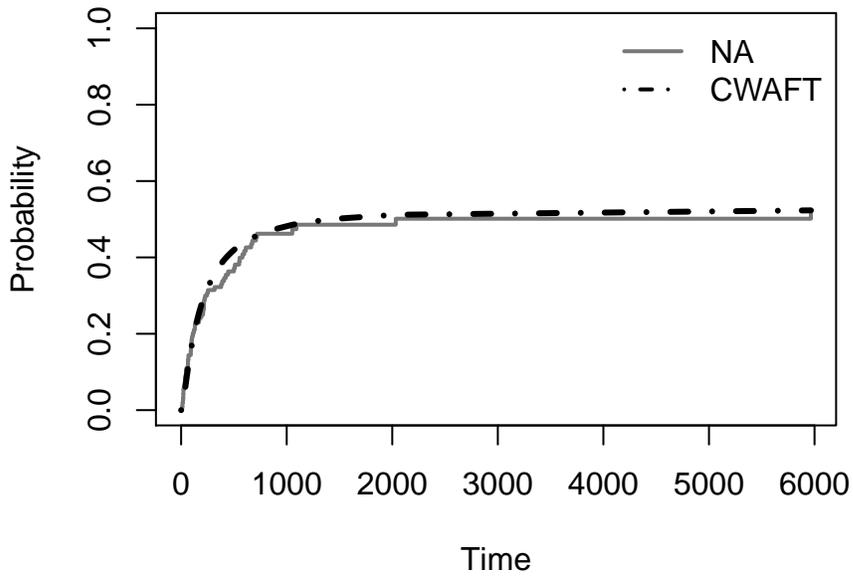}
\caption{Estimated cumulative incidence functions overlaid on the Nelson-Aalen estimators for simulation 2 for groups 1 (top sub-figure) and 2 (bottom sub-figure), respectively.}
\label{sim2cuminc}
\end{figure}

The overall survival function and cumulative incidence curves are shown in Figures \ref{sim2osf} and \ref{sim2cuminc}. The fitted overall survival function and cumulative incidence curves capture the trend of the non-parameteric estimators very well. Parameter estimates and their respective standard errors are presented in Table \ref{parestsim2} for the covariates. The estimated parameter values agree quite closely with the true parameter values. Note that comparison of the true regression parameters ($b_0$, $b_1$, $\sigma^2_g$) with the estimated values is not informative due to censoring.

\subsection{Real Data}
\label{real}
\subsubsection{Stanford heart data}
\label{stanhd}
The Stanford heart dataset is a classical dataset first introduced in \cite{crowley1977}. The data consist of measurements on patients who underwent transplant surgery. We consider the subset of 65 patients with no missing values for the covariates of interest. The failure times recorded were for deaths (in days) from one of two attributed causes: transplant rejection or other causes. This subset contained 29 rejection deaths and 12 deaths from other causes. Twenty-four censored observations did not have an associated failure cause. Information was also obtained on three continuous covariates of interest:  age at the time of transplant, waiting time from acceptance into the program until the day of the surgery, and a mismatch score that was lower for better donor-recipient tissue compatibility. Age and mismatch score were transformed to have zero mean and unit variance while waiting time (highly skewed) was log transformed to make its distribution more similar to a normal distribution.

A criteria such as the Akaike information criterion (AIC) \citep{akaike1974} or the Bayesian information criterion (BIC) \citep{schwarz1978} can be used to compare between models with different number of covariates. We use a combination of the AIC, BIC, and graphical fits to the nonparametric overall survival function and cumulative incidence functions to select the final model. Models with age as the only covariate (model A; AIC=446.79, BIC=470.71) and with mismatch score as the only covariate (model B; AIC=460.49, BIC=484.41) seem to fit quite well. Table \ref{pareststanford} contains the parameter estimates and standard errors for these two models. These data have been analyzed extensively, most famously in \cite{larson1985} who picked a model with both mismatch score and age as the covariates. Our model fit with both mismatch score and age performs similarly to the model with age in terms of visual fit (results not shown), but results in inferior AIC and BIC values.

\begin{table}
\caption{Parameter estimates and standard errors (rounded off to 2 decimals) for models A and B.} \label{pareststanford}
\begin{tabular}{p{0.8in}p{1.2in}p{1.2in}p{1.2in}p{1.2in}}
\hline
Parameter & Model A ($g$=1) & Model A ($g$=2) &  Model B ($g$=1) & Model B ($g$=2)\\
  \hline
$\pi_g$ & 0.65 (0.04) & 0.35 (0.04) & 0.77 (0.03) & 0.23 (0.03)\\ 
$\bmu_g$ & 0.37 (0.1) & -0.68 (0.26) & 0.08 (0.13) & -0.27 (0.35)\\
$\bSigma_g$  & 0.35 (0.09) & 1.45 (0.31) & 0.76 (0.15) &1.6 (0.73)\\
$b_{0g}$ & 5.83 (0.38) & 4.54 (0.21) & 5.83 (0.20) &3.77 (0.29)\\
$\bb_{g}$ & -0.84 (0.75) & -0.88 (0.19) &-0.67 (0.49) &-0.12 (0.28)\\
$\sigma^2_{g}$ & 2.41 (0.38) & 8.93 (0.23) &2.58 (0.37) &7.41 (0.22)\\
   \hline
\end{tabular}

Parameter estimates with standard errors in parenthesis for models A and B, where $g$=1 and 2 refer to the cause of interest and the competing cause, respectively.
\end{table}

\begin{figure}[!h]
\includegraphics{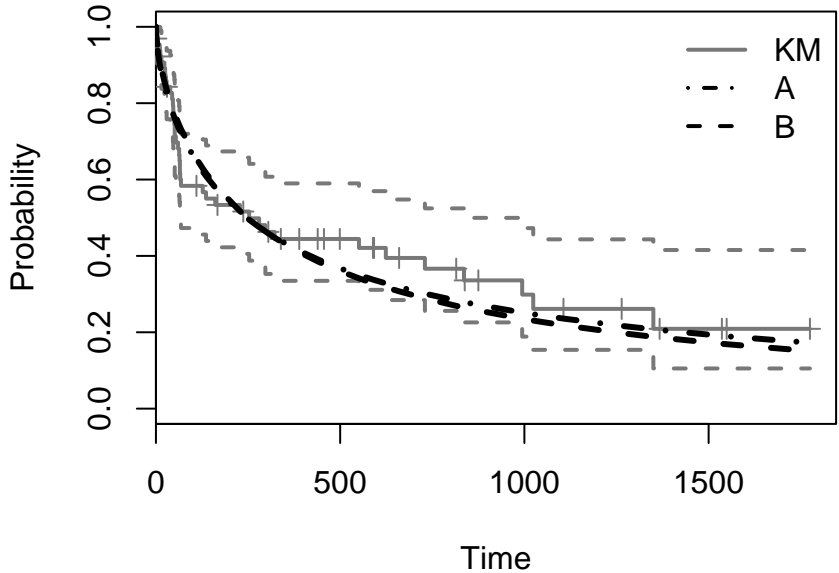}
\caption{Estimated overall survival function overlaid on the Kaplan-Meier estimator and its associated confidence intervals for the Stanford heart data. Here, models A and B refer to the models with age and mismatch score as the only covariates, respectively.}
\label{real1osf}
\end{figure}

\begin{figure}
\includegraphics{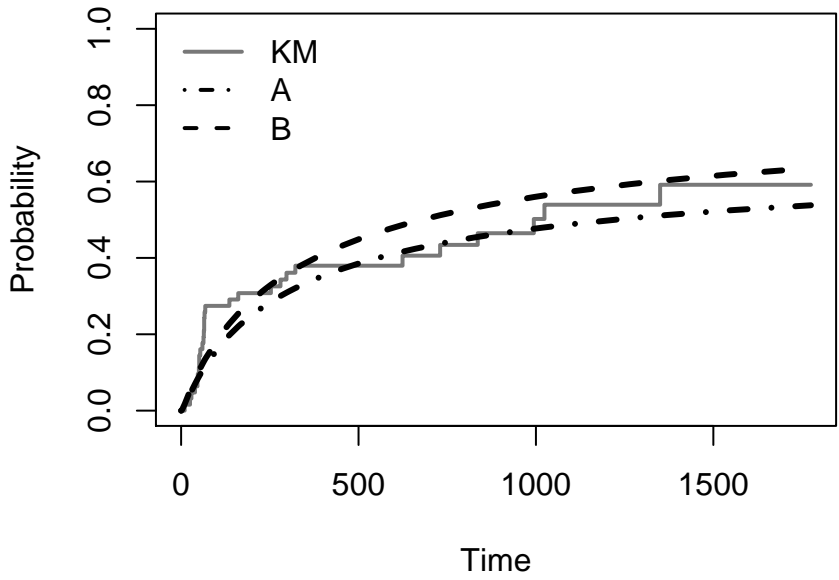}\\
\includegraphics{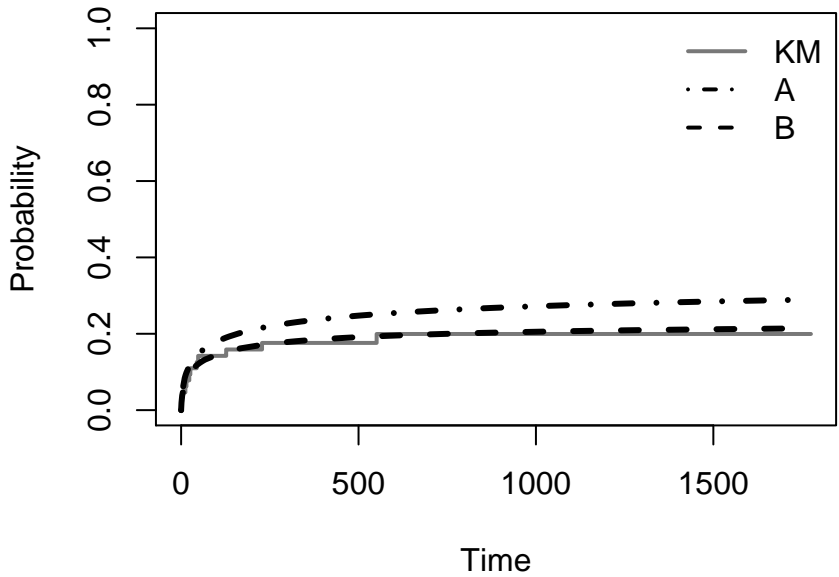}
\caption{Estimated cumulative incidence functions overlaid on the Nelson-Aalen estimators for the Stanford heart data for the cause of interest (top sub-figure) and competing risk (bottom sub-figure), respectively. Models A and B refer to the models with age and mismatch score as the only covariates, respectively.}
\label{real1cuminc}
\end{figure}

The overall survival function and cumulative incidence curves for both models selected are shown in Figures \ref{real1osf} and \ref{real1cuminc}. The fitted overall survival function for both models seems to capture the overall trend of the non-parametric estimators well. Recall that only 12 known deaths were due to the competing event, and the models deal with this differently. The fitted cumulative incidence function for the cause of interest seems to fit better than for the competing event for model A. On the other hand, the reverse is true for model B. The estimated probability of death due to rejection ($\pi_1$) is 0.65 (standard error=0.04) and 0.77 (standard error=0.03) for models A and B, respectively.

\begin{figure}[!h]
\includegraphics{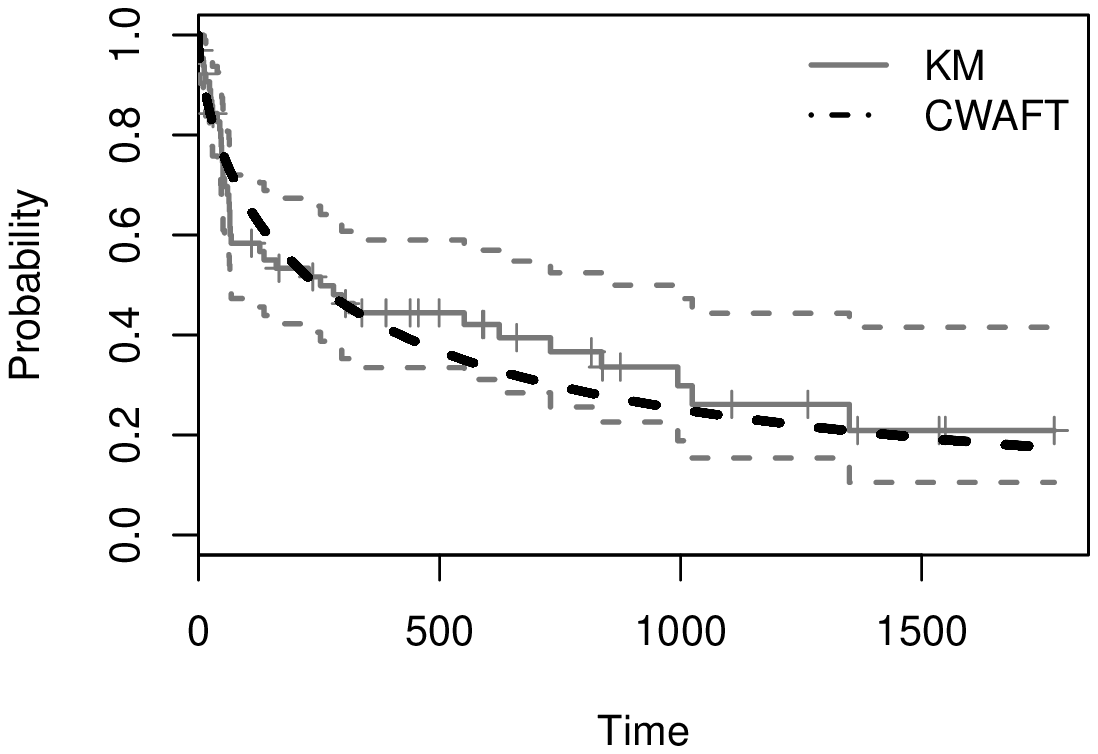}
\caption{Estimated overall survival functions overlaid on the Kaplan-Meier estimator and its associated confidence intervals from 1000 randomly initialized runs for the Stanford heart data.}
\label{heartdata1000}
\end{figure}

The algorithm is also stable from a computational point of view. To demonstrate this, the algorithm was run on the Stanford Health data set one thousand times (with mismatch score as the only covariate) with different initializations for the EM algorithm; the associated fitted overall survival functions were overlaid on the KM estimator curve (Figure \ref{heartdata1000}). 

\subsubsection{4D}
\label{4D}
The 4D data were obtained from \cite{etm2011} and consist of control group data that were assigned to placebo treatment in the 4D study \citep{wanner2005}. The original study was a randomized controlled trial aimed at comparing atorvastatin to placebo for patients with type 2 diabetes and receiving hemodialysis. Patients were followed until death, loss to follow-up, or end of study. The event of interest was a composite of death from cardiac causes, stroke, and non-fatal myocardial infarction; the competing event was death from all other causes. For the subset of female subjects, there were 119 failures observed due to the event of interest and 58 due to the competing event; data from 115 (39\%) participants were censored. Information was also obtained on the age of the patient and transformed to have zero mean and unit variance. Time is measured in years. 

\begin{table}
\caption{Parameter estimates and standard errors (rounded off to 2 decimals) for 4D data.} \label{parest4D}
\begin{tabular}{p{0.8in}p{1.2in}p{1.2in}p{1.2in}p{1.2in}}
\hline
Parameter & Estimates (g=1) & Standard error (g=1) & Estimates (g=2) & Standard error (g=2)\\
  \hline
$\pi_g$ & 0.70 & 0.01 & 0.30  & 0.01\\ 
$\bmu_g$ & -0.11 & 0.08 & 0.27 & 0.10\\
$\bSigma_g$  & 1.09 & 0.14 & 0.67& 0.10\\
$b_{0g}$ & 0.98 & 0.02 & 0.80 & 0.04\\
$\bb_{g}$ & -0.07 & 0.07 & 0.13  & 0.15\\
$\sigma^2_{g}$ & 1.31 & 0.02 & 0.92 & 0.04\\
   \hline
\end{tabular}

Parameter estimates with standard errors in parenthesis, where $g$=1 and 2 refer to the cause of interest and the competing cause, respectively.
\end{table}

\begin{figure}[!h]
\includegraphics{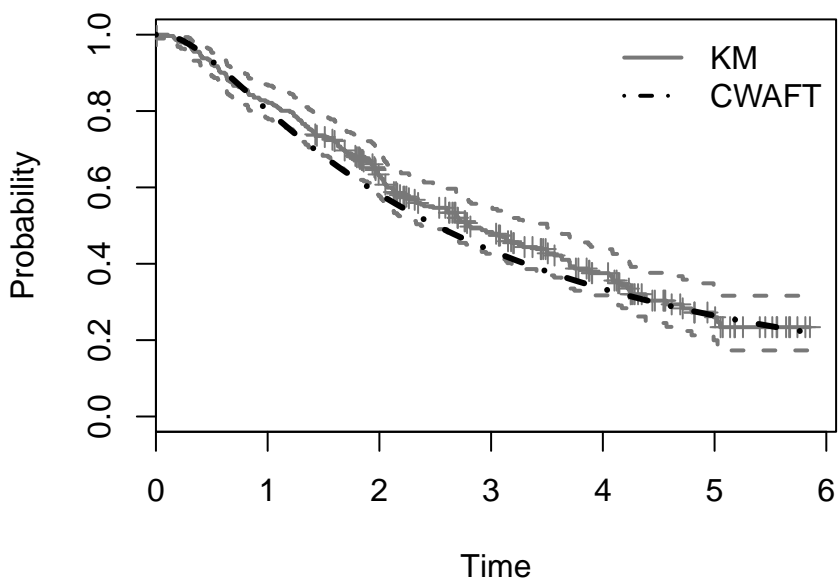}
\caption{Estimated overall survival function overlaid on the Kaplan-Meier estimator and its associated confidence intervals for the 4D data.}
\label{real2osf}
\end{figure}

\begin{figure}
\includegraphics{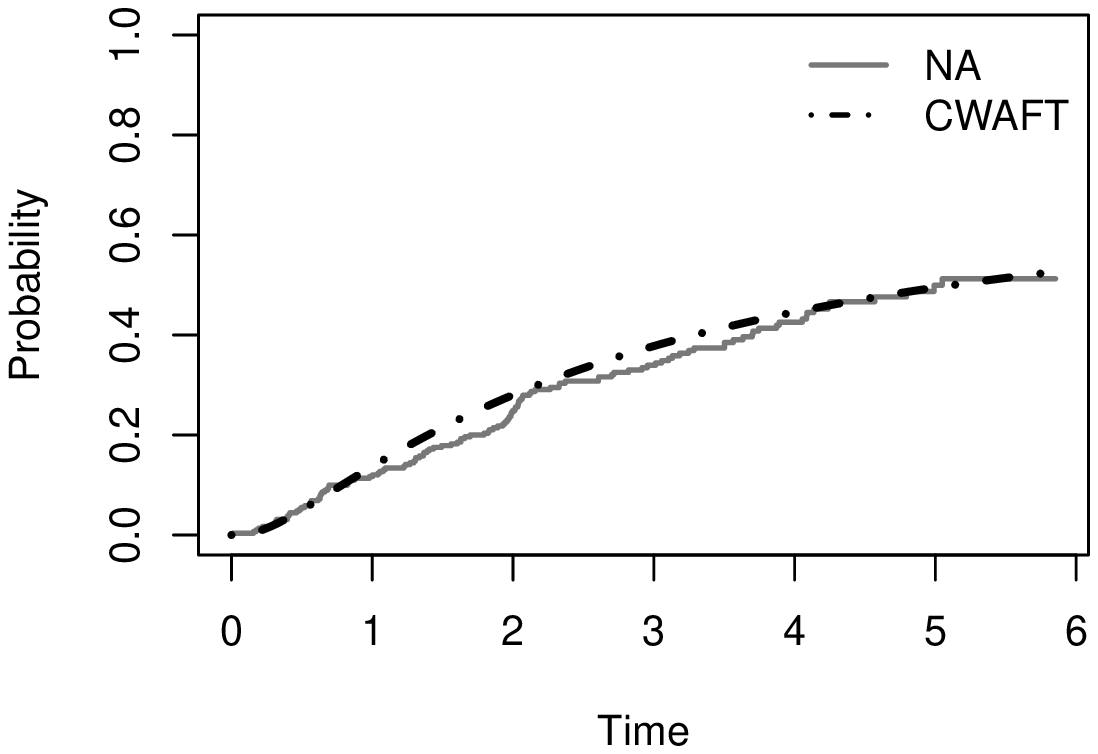}\\
\includegraphics{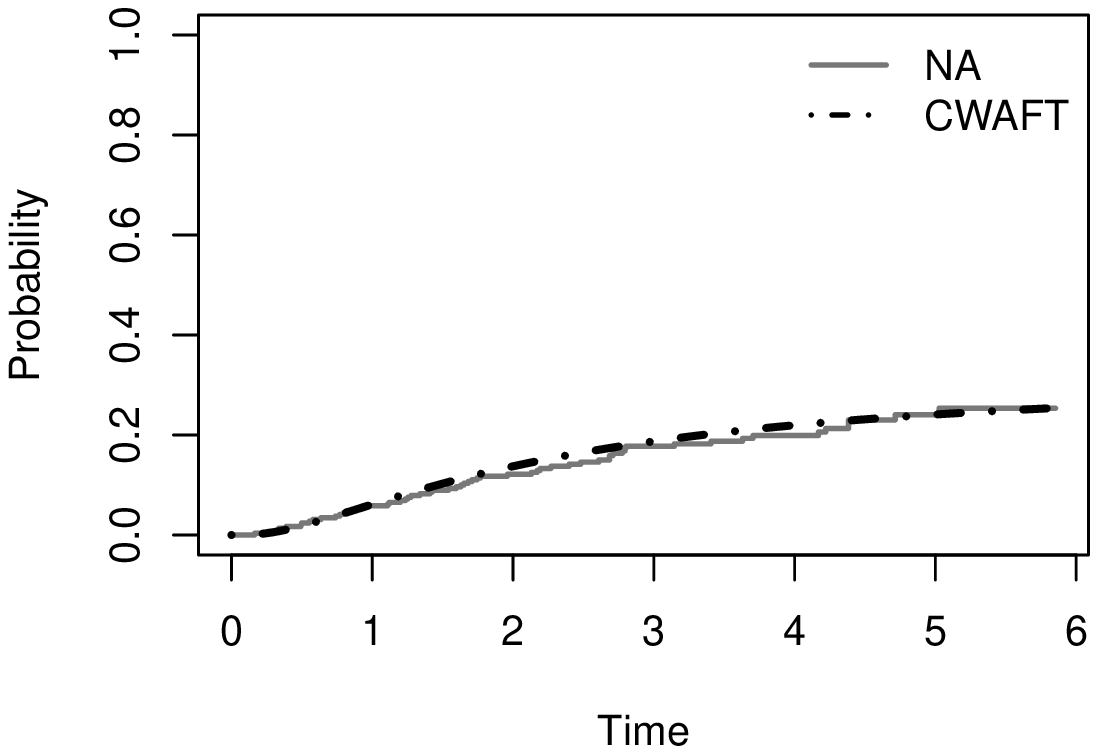}
\caption{Estimated cumulative incidence functions overlaid on the Nelson-Aalen estimators for the 4D data for the cause of interest (top sub-figure) and competing risk (bottom sub-figure), respectively.}
\label{real2cuminc}
\end{figure}

Table \ref{parest4D} contains the parameter estimates and standard errors. Apart from the high percentage of censoring, this dataset is also unique in that censoring begins relatively late (Figure \ref{real2osf}). The estimated cumulative incidence functions are still able to capture the trend of the non-parameteric estimators quite closely (Figure \ref{real2cuminc}). 

\section{Discussion}
\label{discussion}
A novel approach to classifying censored data from different groups while utilizing the distribution of the covariates was presented. In simulated and real data sets, the algorithm showed good performance and was able to extract marginal probabilistic behaviour quite well. The algorithm is quite stable as shown in regards to initialization for the EM. Sometimes, a reasonable number was substituted where NaN or infinity errors were encountered while calculating $\EE^{(k)}(y_{ig})$ or $\EE^{(k)}(y_{ig}^2)$.

Commonly, in the literature, the distribution of the covariates is taken into account to alleviate the impact of any mis-recorded covariates, unobserved heterogeneity, or in the presence of evidence that the data are from particular subsamples of the population. These usually correspond to data misspecification (or missing data) models, or random effect models. In survival literature, frailty models attempt to account for unobserved heterogeneity by including a random effect. Here, the marginal distribution of the covariates $f(\bx)$ is directly taken into account in the maximization of the likelihood function in the EM algorithm. This is also novel in the field of survival analysis to the authors' knowledge.

Note that for both real data sets, the competing risk was death from other events. Also note that it is not always easy to distinguish between a cured individual and a susceptible individual with a large failure time. \cite{farewell1986} gives a good discussion on how care must be exercised in assuming that a fraction of patients is cured. Having said that, a cure rate can estimated by following \cite{ng1998}. Assuming that a patient is cured if the patient dies from a competing cause at a time $T_0$ (for example, $T_0=5$ years has been used in breast cancer studies) without any symptoms of the disease, then from our model, the cure rate can be calculated as $\pi_2S_2(T_0)$. Here, $\pi_2$ is the proportion of individuals who fail from one of the competing causes, that is, will not fail from the cause of interest. $S_2$ refers to the conditional survival function for failure from a competing cause. Note that here $\pi_2$ is being adjusted by excluding those patients with death times smaller than $T_0$ as being cured to account for early deaths that may have been due to postoperative complications and not because they did not have any symptoms of the cause of interest.

Because we are fitting a Gaussian distribution to the marginal distribution of the covariates, currently only continuous covariates can be used. A convenient way to include discrete covariates might be to use them as concomitant variables in a logistic model of the covariates for the probability of failure from a particular risk. However, more work is required to eliminate this aspect from the methodology. Incorporating something akin to a latent trait model for the covariates might be fruitful, for instance if we have data on a few levels of dosage. We used the log-normal distribution here, but other commonly used distributions like log-logistic, Weibull, generalized gamma, etc.\ could also be used depending on the data. Furthermore, semi-parametric extensions of the model might lead to more flexible alternatives. Due to the nature of the likelihood formulation, dealing with missing data in the covariates might also be worthwhile.

\section*{Acknowledgements}
This work is supported by a Alexander Graham Bell Canada Graduate Scholarship (CGS-D) and Discovery Grant from the Natural Sciences and Engineering Research Council (NSERC) of Canada.

\section*{Appendix}
\appendix
\label{app}
\section{Derivation of $\EE^{(k)}(y)$}
\label{AppEyderive}
\begin{align*} 
\EE^{(k)}(y) & = \int_{y^*}^{\infty} y\frac{f(y)}{1-F(y^*)}dy, \\
 & = \int_{y^*}^{\infty} \left(\frac{y-\mu}{\sigma}\right)\frac{\frac{1}{\sqrt{2\pi}\sigma}\exp{\left(\frac{-(y-\mu)^2}{2\sigma^2}\right)}}{1-\Phi(y^*)}dy + \int_{y^*}^{\infty} \frac{\mu}{\sigma}\frac{\frac{1}{\sqrt{2\pi}\sigma}\exp{\left(\frac{-(y-\mu)^2}{2\sigma^2}\right)}}{1-\Phi(y^*)}dy. \\
\end{align*}
Let $z=(y-\mu)/\sigma$. Then, $\sigma dz=dy$ and
\begin{align*}
\EE^{(k)}(y) & = \int_{\frac{y^*-\mu}{\sigma}}^{\infty}z\frac{\frac{1}{\sqrt{2\pi}\sigma}\exp{\left(\frac{-z^2}{2}\right)}}{1-\Phi\left(\frac{y^*-\mu}{\sigma}\right)}\sigma dz + \int_{\frac{y^*-\mu}{\sigma}}^{\infty} \frac{\mu}{\sigma}\frac{\frac{1}{\sqrt{2\pi}\sigma}\exp{\left(\frac{-z^2}{2}\right)}}{1-\Phi\left(\frac{y^*-\mu}{\sigma}\right)} \sigma dz, \\
& = \frac{1}{1-\Phi\left(\frac{y^*-\mu}{\sigma}\right)}\int_{\frac{y^*-\mu}{\sigma}}^{\infty} \sigma z \frac{1}{\sqrt{2\pi}}\exp{\left(\frac{-z^2}{2}\right)} dz +  \\
& \qquad \frac{1}{1-\Phi\left(\frac{y^*-\mu}{\sigma}\right)} \int_{\frac{y^*-\mu}{\sigma}}^{\infty} \mu\frac{1}{\sqrt{2\pi}}\exp{\left(\frac{-z^2}{2}\right)} dz , \\
& = \frac{1}{1-\Phi\left(\frac{y^*-\mu}{\sigma}\right)}\left[\sigma \phi\left(\frac{y^*-\mu}{\sigma}\right)  +  \mu \left(1-\Phi\left(\frac{y^*-\mu}{\sigma}\right)\right) \right], \\
& = \mu + \sigma \frac{\phi\left(\frac{y^*-\mu}{\sigma}\right)}{1-\Phi\left(\frac{y^*-\mu}{\sigma}\right)}. \\
\end{align*}

\section{Derivation of $\EE^{(k)}(y^2)$}
\label{AppEy2derive}
\begin{align*} 
\EE^{(k)}\left(y^2\right) & = \int_{y^*}^{\infty} y^2\frac{f(y)}{1-F(y^*)}dy, \\
& = \int_{y^*}^{\infty} y^2\frac{\frac{1}{\sqrt{2\pi}\sigma}\exp{\left(\frac{-(y-\mu)^2}{2\sigma^2}\right)}}{1-\Phi\left(y^*\right)}dy, \\
& = \int_{y^*}^{\infty}\sigma \left(\frac{y-\mu}{\sigma}\right)^2\frac{\frac{1}{\sqrt{2\pi}}\exp{\left(\frac{-(y-\mu)^2}{2\sigma^2}\right)}}{1-\Phi\left(y^*\right)}dy + \\
& \qquad  \int_{y^*}^{\infty} \sigma\left(\frac{2y\mu-\mu^2}{\sigma^2}\right)\frac{\frac{1}{\sqrt{2\pi}}\exp{\left(\frac{-(y-\mu)^2}{2\sigma^2}\right)}}{1-\Phi\left(y^*\right)}dy, \\
& = \int_{y^*}^{\infty} b \sigma \left(\frac{y-\mu}{\sigma}\right)^2\frac{1}{\sqrt{2\pi}}\exp{\left(\frac{-(y-\mu)^2}{2\sigma^2}\right)}dy + \\
& \qquad 2b\mu\EE^{(k)}(y)(1-\Phi\left(y^*\right)) + \\
& \qquad \int_{y^*}^{\infty} b \sigma\left(\frac{-\mu^2}{\sigma^2}\right)\frac{1}{\sqrt{2\pi}}\exp{\left(\frac{-(y-\mu)^2}{2\sigma^2}\right)}dy, \\
\end{align*}
where $b=1/\left(1-\Phi\left(y^*\right)\right)$. Again, let $z=(y-\mu)/\sigma$. Hence,
\begin{align*} 
\EE^{(k)}\left(y^2\right) & = \int_{\frac{y^*-\mu}{\sigma}}^{\infty} b^* \sigma z^2 \frac{1}{\sqrt{2\pi}}\exp{\left(\frac{-z^2}{2}\right)}\sigma dz + \\
& \qquad 2 b^* \mu\EE^{(k)}(y)\left(1-\Phi\left(\frac{y^*-\mu}{\sigma}\right)\right) - b^* \mu^2\left(1-\Phi\left(\frac{y^*-\mu}{\sigma}\right)\right) , \\
\end{align*}
where $b^*=1/\left(1-\Phi\left(\frac{y^*-\mu}{\sigma}\right)\right)$. Now, using integration by parts, 

\begin{align*}
\EE^{(k)}\left(y^2\right) & = b^*\sigma^2\left[\left(\frac{y^*-\mu}{\sigma}\right)\phi{\left(\frac{y^*-\mu}{\sigma}\right)} + 1 - \Phi\left(\frac{y^*-\mu}{\sigma}\right)\right] +   \\
& \qquad 2 b^* \mu\EE^{(k)}(y)\left(1-\Phi\left(\frac{y^*-\mu}{\sigma}\right)\right) -b^*\mu^2\left(1-\Phi\left(\frac{y^*-\mu}{\sigma}\right)\right), \\
 & = \sigma_g^2\left[\frac{\left(\frac{y^*-\mu_g}{\sigma_g}\right)\phi{\left(\frac{y^*-\mu_g}{\sigma_g}\right)} + 1 - \Phi\left(\frac{y^*-\mu_g}{\sigma_g}\right)}{1-\Phi\left(\frac{y^*-\mu_g}{\sigma_g}\right)}\right] + 2\mu\EE^{(k)}(y_{ig}) -\mu^{2}_{g} . \\
\end{align*}

\section{Maximization step}
\label{mstepapp}
For $\hat{b}_{0g}^{(k+1)}$ $(g=1, \ldots,G)$, $\sum_{i=1}^N \frac{\partial \left(\tau_{ig}^{(k)} Q_{1}\left(\bchi_g|\bpsi^{(k)}\right)\right)} {\partial b_{0}}=0$
yields
\begin{align*}
\sum_{i=1}^N \tau_{ig}^{(k)} \left[\EE^{(k)}(y_{ig}) - \left(\bb_g^{'(k)}\bx_i + b_{0g}^{(k)}\right)\right] &= 0 \\
\sum_{i=1}^N \tau_{ig}^{(k)} \left(\EE^{(k)}(y_{ig})-\bb_g^{'(k)}\bx_i\right) &=  b_{0g}^{(k)} \sum_{i=1}^N \tau_{ig}^{(k)},
\end{align*}
and then we get
\begin{equation*}
\hat{b}_{0g}^{(k+1)} = \frac{\sum_{i=1}^N \tau_{ig}^{(k)} \EE^{(k)}(y_{ig})}{\sum_{i=1}^N \tau_{ig}^{(k)}}-\bb_{g}^{'(k+1)}\frac{\sum_{i=1}^N \tau_{ig}^{(k)} \bx_i}{\sum_{i=1}^N \tau_{ig}^{(k)}}. 
\end{equation*}
Similarly, for $\hat{\bb}_{g}^{(k+1)}$ $(g=1, \ldots,G)$, $\sum_{i=1}^N \frac{\partial \left(\tau_{ig}^{(k)}  Q_{1}\left(\bchi_g;|\bpsi^{(k)}\right)\right)} {\partial \bb'_{g}}  = \boldsymbol{0}'$ implies
\begin{equation*}
\sum_{i=1}^N \tau_{ig}^{(k)} \left[\EE^{(k)}(y_{ig})-\left(\bb_g'\bx_i+b_{0g}^{(k)}\right)\right]\bx'_n  = \boldsymbol{0}' ,
\end{equation*}
yielding
\begin{align*}
\hat{\bb}_{g}^{'(k+1)} &= \left(\frac{\sum_{i=1}^N \tau_{ig}^{(k)} \EE^{(k)}(y_{ig}) \bx'_n}{\sum_{i=1}^N \tau_{ig}^{(k)}}-\frac{\sum_{i=1}^N \tau_{ig}^{(k)} \EE^{(k)}(y_{ig})}{\sum_{i=1}^N \tau_{ig}^{(k)}}\frac{\sum_{i=1}^N \tau_{ig}^{(k)} \bx'_n}{\sum_{i=1}^N \tau_{ig}^{(k)}}\right) \times  \\  
& \qquad \left(\frac{\sum_{i=1}^N \tau_{ig}^{(k)} \bx_i\bx'_n}{\sum_{i=1}^N \tau_{ig}^{(k)}}-\frac{\sum_{i=1}^N \tau_{ig}^{(k)} \bx_i}{\sum_{i=1}^N \tau_{ig}^{(k)}} \left(\frac{\sum_{i=1}^N \tau_{ig}^{(k)} \bx'_n}{\sum_{i=1}^N \tau_{ig}^{(k)}}\right)  \right)^{-1}. \\  
\end{align*}
For $\hat{\sigma}_{g}^{(k)}$ $(g=1, \ldots,G$), $\sum_{i=1}^N \frac{\partial\left(\tau_{ig}^{(k)}  Q_{1}(\bchi_g|\bpsi^{(k)})\right)} {\partial \sigma_{g}^{2}}=0$ yields 
\begin{equation*}
\sum_{i=1}^N \tau_{ig}^{(k)} \left( -\frac{1}{\sigma_{g}^{2(k)}} + \frac{1}{\sigma_{g}^{4(k)}} \EE^{(k)}\left(y_i-\left(\bb_g^{'(k)}\bx_i+b_{0g}^{(k)}\right)\right)^2 \right) = 0.
\end{equation*}
This implies
\begin{equation*}
\hat{\sigma}_{g}^{2(k+1)} = \frac{\sum_{i=1}^N \tau_{ig}^{(k)} \EE^{(k)}\left(y_i-\left(\bb_g^{'(k+1)}\bx_i+b_{0g}^{(k+1)}\right)\right)^2} {\sum_{i=1}^N \tau_{ig}^{(k)}} \label{Msigma(k+1)}.  
\end{equation*}

\end{document}